\documentclass[12pt, A4paper]{article}
\usepackage{caption}
\usepackage{subcaption}
\usepackage{amsmath}
\usepackage{verbatim}
\usepackage{setspace}
\usepackage{rotating}
\usepackage{morefloats}
\usepackage{graphicx, xcolor}
\usepackage{color, colortbl}
\usepackage{url}
\usepackage{amssymb}
\usepackage[sectionbib]{natbib}
\usepackage{hyperref}
\usepackage{sectsty}
\usepackage{multirow}
\usepackage{longtable}
\usepackage{rotating}
\usepackage[labelfont=bf,font=footnotesize]{caption}
\usepackage[left=2.56cm,right=2.56cm,top=2.56cm]{geometry}
\usepackage{mathrsfs}
\usepackage{dsfont}
\usepackage{enumerate,letltxmacro}
\usepackage[normalem]{ulem}
\LetLtxMacro\itemold\item
\newcommand{\bt}{\pmb{\theta}}

\newcommand{\bh}{\pmb{\text{h}}}

\definecolor{Gray}{gray}{0.9}

\begin{document}
\clearpage
\setcounter{page}{1}
\title{Triangulating War: Network Structure and the Democratic Peace}
\author{Benjamin W. Campbell\footnote{BWC: Doctoral Candidate in Political Science, The Ohio State University, e: \href{mailto:campbell.1721@osu.edu}{campbell.1721@osu.edu}} \and Skyler J. Cranmer\footnote{SJC: Carter Phillips and Sue Henry Associate Professor of Political Science, The Ohio State University, e: \href{mailto:cranmer.12@osu.edu}{cranmer.12@osu.edu}} \and Bruce A. Desmarais\footnote{BAD: Associate Professor of Political Science, The Pennsylvania State University, e: \href{mailto:bdesmarais@psu.edu}{bdesmarais@psu.edu}}}

\maketitle
%\tableofcontents
\begin{center} \vspace{-1cm} Word Count: 9,884 \end{center}
\begin{abstract}

\noindent Decades of research has found that democratic dyads rarely exhibit violent tendencies, making the democratic peace arguably the principal finding of Peace Science.
However, the democratic peace rests upon a dyadic understanding of conflict.  Conflict rarely reflects a purely dyadic phenomena---even if a conflict is not multi-party, multiple states may be engaged in distinct disputes with the same enemy.
We postulate a network theory of conflict that
treats the democratic peace as a function of the competing interests of mixed-regime dyads and the strategic inefficiencies of fighting with enemies' enemies.
Specifically, we find that a state's decision to engage in conflict with a target state is conditioned by the other states in which the target state is in conflict.
When accounting for this network effect, we are unable to find support for the democratic peace.
This suggests that the major finding of three decades worth of conflict research is spurious. \\
\thispagestyle{empty}

\end{abstract}

% Paper Word Count:

\clearpage

\section{Introduction}
Given its societal costs, few phenomena warrant more attention than interstate conflict.  The causes of militarized interstate disputes (MIDs) have been considered for decades, and their study is central to Peace Science.  From this tradition has emerged the democratic peace, the hallmark finding holding that democracies do not fight one another.  Since the early 1990s, this literature has steadily grown into a major body of work, consisting of hundreds of peer-reviewed articles.  The consistent robustness of the democratic peace has lead many to refer to it as a law of international politics \citep{hegre2014democracy}.

In most cases, scholars of the democratic peace consider dyadic conflicts. This focus on bilateral conflict poses a theoretical problem, excluding the complex dependencies at work outside the focal dyad that influences conflict behavior. The dyadic approach carries with it a methodological problem in that regression methods treat dyadic conflicts as independent conditional on the covariates \citep{cranmer2016critique}.  This is puzzling as conflict is an intrinsically relational behavior that often includes more than two parties, even if those parties are not bilaterally connected. For illustration, consider that US action in Syria is constrained both by America's relations with it's allies---particularly Turkey---and geopolitical opponents---particularly Russia.
% This dyadic focus has produced a number of problems.  For an illuminating case, this dyadic focus treats World War I and World War II as 567 \textit{entirely independent} dyadic conflicts, making up $46.14\%$ of conflicts fought since 1816 \textcolor{red}{Cite Cranmer Desmarais and Menninga CMPS}.

We probe \textit{which interdependencies govern the decision of states to engage in militarized conflict} and suggest that \emph{the democratic peace is an artifact of studying conflict from a dyadic perspective.}  We argue that the direct or indirect coordination of states with similar regime types against common enemies accounts for the seemingly robust finding that jointly democratic (and occasionally jointly autocratic) dyads, are less dispute-prone.  To test our theory, we utilize inferential network analysis to uncover the meso-level dynamics that inform state behavior.  We find robust support for our proposition: once we account for the common enemies of jointly democratic or jointly autocratic pairs, statistical support for the democratic (and autocratic) peace vanishes. Moreover, after controlling for common bilateral predictors of conflict and the meso-level network measures we introduce here, jointly democratic dyads appear to be \emph{more} conflict-prone that mixed-regime dyads.

\section{The Democratic Peace}
The democratic peace is one of the foundational findings in IR, acquiring a law-like status for some \citep{hegre2014democracy}.  This finding has been surprisingly robust, and can be reduced as follows: jointly democratic states do not go to war, but democratic states are not monadically less likely to engage in conflict.  Traditionally, scholars have attempted to make sense of this finding through structural or normative explanations \citep{maoz1993normative}.  Alternative explanations rely less upon regime type and more upon many related factors, including trade networks \citep{dorussen2010trade}, economic openness \citep{gartzke2007capitalist}, lootability \citep{rosecrance1986rise}, IGOs \citep{pevehouse2006democratic, dorussen2008intergovernmental}, and identity \citep{gartzke2013permanent}.

A fitting review of every explanation for the democratic peace is beyond the scope of a single book, much less an article.  As such, we will confine our discussion to the two most prominent explanations, the institutional and normative accounts, as well as the most pertinent explanation to our discussion---the network explanation.  While many have challenged the democratic peace using conventional large-N regression analyses, a number of recent papers have used the tools of inferential network analysis to challenge its statistical validity \citep{ward2007disputes,dorussen2008intergovernmental, maoz2009effects, maoz2010networks}---even drawing upon network theory to question its fundamental premise \citep{cranmer2011inferential}.  Given the complex interdependencies established to exist in the conflict network, a fully developed network theory tested using appropriate inferential network analysis methods seems overdue.

\subsection{Institutional Explanations}
The initial explanation for the democratic peace holds that the structural and institutional constraints placed on democratic policymakers lead to peaceful behavior when interacting with other democracies.  Introduced by \citet{rummel1983libertarianism}, this theory rests upon two assumptions \citep{maoz1993normative}.  First, in order to retain the benefits of holding office,  leaders must mobilize selectorate support for their policies.  Second, mobilization of this necessary political support can only be shortcut in times of emergency. Given these propositions, democratic leaders are reluctant to wage war given the cost, and should they decide to go to war, preparation requires more time.  As a result, by the time two democracies are mobilized for war, diplomats would have had the opportunity to resolve disputes peacefully \citep{maoz1993normative, russett1994grasping}.%\footnote{There are a variety of other ``institutional" explanations for the democratic peace, however, most of these theories are monadic explanations.  For example, a monadic theory argues that democratic leaders face institutional constraints (e.g. public opinion), making it increasingly difficult for them to go to war \citep{immanuel1991perpetual}.}  
Alternatively, when democracies experience crises with autocrats, who lack democratic constraints on mobilization and escalation, they must act quickly -- they are put in no-choice situations \citep{rummel1983libertarianism, maoz1993normative, russett1994grasping}.

Another form of the institutional explanation emphasizes audience costs. In democracies, leaders incur costs for backing down or reverting from stated positions.  These costs may make it easier for democracies to credibly signal commitments and resolve disputes \citep{fearon1994domestic, levy2015backing}.  In other words, democracies have the ability to signal resolve, and in jointly democratic dyads, they are capable of more easily resolving disputes as both sides are capable of credibly committing to a negotiated settlement \citep{hegre2014democracy}.   It is worth noting, however, that the audience cost literature is not without its critiques, with many saying that there is little empirical support for audience costs \citep{snyder2011cost, downes2012illusion}, and that their logic should extend to single-party autocratic regimes were it true \citep{weeks2008autocratic}.

\subsection{Normative Explanations}
Another classic and widely valued explanation of the democratic peace holds that democracies act in accordance with certain norms that lead to peace when interacting with other democracies.  This explanation, developed by \citet{doyle1986liberalism}, typically makes two assumptions \citep{maoz1993normative, russett1994grasping}.  First, states (democratic or autocratic), to the extent possible, externalize norms of behavior that are developed internally and characterize their domestic political processes and institutions.   Second, the international system lacks a centralized authority determining norms, and when democratic and non-democratic norms clash in dyadic relations, non-democratic norms will be dominant. This leads to an important theoretical proposition, best described by \citet{ember1992peace}: `The culture, perceptions, and practices that permit compromise and the peaceful resolution of conflicts without the threat of violence within countries come to apply across national boundaries towards other democratic countries as well," (576).
%There have been a variety of other implications drawn from the normative model.   \citet{dixon1993democracy} and \citet{leng1993reciprocating} argue that disputes between democracies are more likely to be settled by third-party conflict management, mediation, agreement, or settlement.\footnote{This might be considered an extension of the institutional account as many of these accounts rely upon information transparency and commitment credibility, however proponents argue that norms lead states to come to the table in the first place.} %An additional explanation holds that the perception of instability within democracies may lead states to treat democracies as non-democracies, and many of the norms governing behavior between democracies may not be present  \citep{huth1993general, maoz1992alliance, maoz1989joining}.
Many dispute the normative approach to understanding the democratic peace by arguing that its theories should predict a monadic democratic peace that does not exist \citep{raknerud1997hazard}, and that democracies often break liberal norms  \citep{rosato2003flawed}.

%\begin{itemize}
%\item \citet{doyle1986liberalism} represented the normative account of the democratic peace very initially, building upon \citet{immanuel1991perpetual} and Woodrow Wilson.  There were two assumptions made for this explanation.  First, states, to the extent possible, externalize norms of behavior that are developed within the state and characterize their domestic political processes and institutions.  Second, the anarchic nature of international politics implies that there is some clash between democratic and nondemocratic norms, and the nondemocratic norms dominate the democratic norms.
%\item Disputes between democracies are more likely to be settled by third-party conflict management, agreement, or stalemate \citep{dixon1993democracy, leng1993reciprocating}
%\item Norms can be transmitted to international environments.  Perception of instability may encourage the use force against the unstable regime \citep{huth1993general, maoz1992alliance, maoz1989joining}
%\item \citet{ember1992peace}: ``the culture, perceptions, and practices that permit compromise and the peaceful resolution of conflicts without the threat of violence within countries come to apply across national boundaries towards other democratic countries as well," (576).
%\item There are obviously responses -- this should be monadic \citet{raknerud1997hazard}, and liberal states often break norms \citet{rosato2003flawed}.
%\end{itemize}

\subsection{Network Explanations}
Scholars have recently considered the democratic peace through a networks-based framework, emphasizing the explanatory power of trade \citep{maoz2006structural, dorussen2010trade, lupu2012trading}, IGOs networks \citep{maoz2006structural, dorussen2008intergovernmental}, threat-based ego-networks \citep{maoz2010networks} and interdependence broadly \citep{maoz2009effects, maoz2010networks}.  To date, few have attempted to understand the democratic peace using the topology---the structure of network connections---of the conflict network \citep{maoz2007enemy, cranmer2011inferential}.

The trade network has been used to explain the democratic peace.  \citet{maoz2006structural} found that similarity in states' structural positions in the trade network dampens the probability of dyadic conflict.  \citet{dorussen2010trade} and \citet{lupu2012trading} both find evidence for the direct and higher-order pacifying effects of trade.  Similar to their findings on trade, \citet{maoz2006structural} find pacifying effects for structural similarity in the IGO network.  \citet{dorussen2008intergovernmental} finds that indirect links between states though IGO networks decrease the probability of conflict, particularly when diplomatic ties are weak.

Maoz posits that network explanations may resolve the \textit{democratic peace paradox} wherein the democratic peace holds at the dyadic level, but not at the monadic or systemic levels.  \citet{maoz2010networks} argues that as the number of democracies within a state's threat-based ego-network increases, a democracy is able to behave according to democratic norms and not revert to the realist norms of autocracies.  One problem with this approach is that it seems much more likely that states would adopt particular norms, democratic and liberal or autocratic and realist, according to the particular state they are interacting with than the composition of the broader ego-network.
%In his 2010 book, Maoz posits that network explanations may offer a means to resolve the \textit{democratic peace paradox} wherein the democratic peace holds at the dyadic level, but not at the monadic or systemic levels.  \citet{maoz2010networks} argues that as the number of democracies within a state's threat-based ego-network increases, a democracy is able to behave according to democratic norms and not revert to the realist norms of autocracies.  %Maoz claims that this logic explains the monadic and dyadic democratic peace, and may indicate when and where more democratic systems should produce peace.  While this may be considered a network theory, at its core it relies more upon the strategic interaction of states and the role of norms in crisis bargaining than the structure of the conflict network.  This theory, however, assumes that autocracies are not incentivized to exploit their democratic partners and revert to realist norms.
%It seems much more likely that states would adopt particular norms, democratic and liberal or autocratic and realist, according to the particular state they are interacting with than a characteristic of the broader ego-network.

To our knowledge there has been only one, albeit brief, attempt to derive a purely network topology-based explanation for the democratic peace.  \citet{cranmer2011inferential} argue that two features of the conflict network may explain the democratic peace (81). First, a state's ``popularity" in the conflict network, or the degree of targeting they experience, may explain certain dynamics that dyadic covariates may not capture.  In particular, this popularity term or ``two-star" term capturing the number of times two states $i$ and $k$ are engaged in a dispute with the third $j$, would capture organized and internationally coordinated responses to diplomatic and security crises.  Second, \citet{cranmer2011inferential} specify a ``triangle" network statistic, which they argue should reflect the tendency for the conflict network to not form closed triangles of conflict.   In this particular case, it would make very little sense for $i$ and $k$ to go to war as it would undermine their mutual effort against $j$ -- we would expect the enemy of a state's enemy to be its friend, or at least not its enemy.\footnote{In related work, \citet{ward2007disputes} attempt to condition out the effect of network interdependencies, such as sender or receiver effects or triadic closure, but do not attempt to introduce a theoretically-motivated network explanation for the democratic peace.}
% Added footnote to WSC

The approach builds upon the findings of \citet{gowa1995democratic} and \citet{farber1997common} while offering a unique, fresh, and more empirically rigorous take.  Farber and Gowa argue that rates of war between democracies are lower than autocratic or mixed dyads due to the Cold War, which lead to a pattern of alliances and interests that emerged as a response to systemic dynamics.  We attempt to build upon these works, devising a more fully developed network theory of the democratic peace, while providing more rigorous analyses to shed light on how precisely network based dynamics influence the democratic peace.

\section{Network Structure and Conflict}
% Brief introduction about theory and what we can gain
Our network theory positions a state's decision to go to war within the broader network-level dynamics that condition state behavior.  This theory essentially holds that the democratic peace is a function of how states are aligned within a network, and how that alignment produces peace between democracies.  This peace is not a direct function of regime type, but the inefficiencies of fighting enemies' enemies.  In studying the effect of network topology on conflict two processes must be studied: (1) by what processes do states make the decision to engage in a MID, and (2) how do network-level dynamics condition these state decision-making processes? %Each of these questions are considered in the following subsections%On the one hand, our theory may be considered more complicated than existing explanations for the democratic peace because it involves more than just a dyadic pair of states. On the other hand, the theoretical mechanism we propose is much simpler than existing explanations offered in the literature.  %We hope that this networks approach can help to make sense of the democratic peace.

% What decisions do states make when deciding to go to war and how do network level effects condition those decisions
.

\subsection{State decision-making and conflict}
% By what processes do states make the decision to engage in a militarized interstate dispute? Fearon/conventional rationalist explanations
It is generally thought that conflict emerges from disputes between states that have diverging interests and are incapable of agreeing to a peaceful settlement \citep{fearon1995rationalist}.  These problems are often resolved through the use of third party mediators \citep{li2017three} or costly signals \citep{fearon1994domestic}.  While this may be a gross simplification of the literature, we find it a useful assumption. The emergence of conflicting preferences, the nature of these disputes, and their likelihood to escalate are all informed by the regime type of the states in question \citep{gartzke1998kant, gartzke2000preferences, werner2000effects, scott2006toward}.  

% Democracies likely to have conflicting preferences with autocracies
If the opportunity for experiencing low-level disputes is a function of state preferences, the political institutions of states surely matter.  The relationship between mixed-regime dyads and MIDs is well documented \citep{gartzke1998kant, gartzke2000preferences, werner2000effects, scott2006toward}; as institutions and regimes become dissimilar, there is an increase in the opportunity for disagreement about foreign or domestic policy preferences \citep{gartzke1998kant, werner2000effects}.  These disputes more more intractable as they may often emerge from differences in preferences that are integral to the state, the identity of its citizens, or its leadership \citep{doyle1983kant}.  Consider, for example, the opportunity for disagreement between a democracy and an autocracy regarding human rights---Cuba and the United States certainly have experienced a myriad of disputes motivated largely by the dissimilarity of their political regimes and human rights abuses under Castro.   This dispute seems to be as intractable as a dispute can get -- for the United States to achieve its goal and then subsequently remove its sanctions on Cuba, Castro had to have stepped down and accepted liberal institutions.

% These disputes are more likely to escalate to militarized, and in particular, fatally militarized status
These intractable disputes often constitute enduring rivalries that occasionally flare into violent conflict.  Escalation is a function of many sub-processes, including a fear of dissimilar regimes and exogenous political shocks.  \citet{weart1994peace} notes that autocratic states, which may be fearful of liberal intervention, retaliate against threats by initiating a dispute to simultaneously resolve the external threat and resolve internal discontent.  Enduring rivalries, particularly between democracies and autocracies, may be more likely to escalate as a function of exogenous shocks, including changes in relative capabilities \citep{geller1993power}, economic productivity \citep{bennett2000foreign}, or leadership \citep{bennett1997democracy}.  To keep with the previous example, following the start of his administration in January 1981, President Reagan adopted a hard-line policy towards Cuba.  In July the Reagan Administration, in an initial display of resolve, conducted air exercises.  This prompted Castro to organize a large militia force out of fear of potential U.S. military action.%\footnote{This is MID number 2971 in the Correlates of War MIDs dataset.}  
The preceding leads to the following hypothesis:
	\begin{enumerate}
	\item[$H_1$] Mixed dyads are more likely to engage in military conflict.
	\end{enumerate}
% 1981 cite: \citep[420]{mazarr1988semper}

Jointly democratic or autocratic dyads are not too distinct in this regard; certainly states can have intractable disputes regardless of regime type.  Jointly autocratic conflict is well documented, and support for the autocratic peace \citep{raknerud1997hazard, scott2006toward} has recently been questioned \citep{gelpi2008democracy, dafoe2013democratic}.  While MIDs between democracies are certainly less common, there are a non-trivial amount of them.  \textit{How, then, can one square the plethora of disputes between mixed dyads with the relative sparsity of disputes between autocratic and democratic dyads?} We believe that the answer lies in structural features of the conflict network:  that states with common enemies should not themselves be enemies. More specifically, the conflicts between mixed dyads often prompt formal or informal coordination between states of a common regime type, and because of that, jointly autocratic states and jointly democratic states are less likely to fight.

\subsection{Network structure and conflict}
% How shared regimes typically coordinate on militarized interstate disputes OR also are more likely to have common enemies
The preceding logic cannot explain the democratic or autocratic peace independently.  Any two states, within reason, could find themselves having a conflict capable of escalation.  We propose that network dynamics explain why these mixed-regime dyadic conflicts are so common relative to jointly autocratic or democratic conflict, which may in fact be artifacts emerging form the inefficiencies associated with fighting enemies of enemies.

\begin{figure}
\centering
\begin{subfigure}{.5\textwidth}
\hspace{1.95cm}
\includegraphics[width=.5\linewidth]{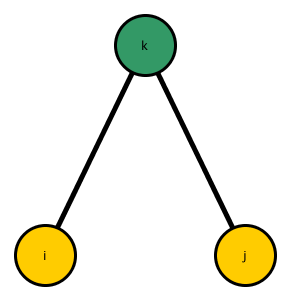}
\caption{Two-star}
\label{2star}
\end{subfigure}%
\begin{subfigure}{.5\textwidth}
\centering
\includegraphics[width=.5\linewidth]{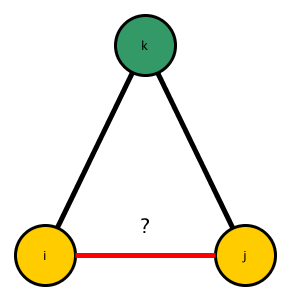}
\caption{Triangle}
\label{triangle}
\end{subfigure}
\caption{\textbf{Generative features of the fatal MID network.}}
\label{subgraphs}
\end{figure}

\citet{cranmer2011inferential} established a generative model of the conflict network using two network processes: \textit{k-stars} and \textit{triangles}.  K-stars, in network parlance, refer to a process where actor $k$ is connected to actors $i$ and $j$, but actors $i$ and $j$ are not connected to each other. %This is referred to as a two-star because $k$ is connected to two actors; if we imagine $k$ having many more connections, the `star' analogy becomes intuitive. 
However, if $k$ is of a different regime type than $i$ and $j$, we refer to this as a mixed two-star because actors included are of different types.  In our case, a mixed two-star reflects a situation in which two states, suppose both democracies or autocracies, are engaged in a MID with a common enemy of the other type.  This situation is illustrated in Figure \ref{2star}.  Triangles refer to the same process but actors $i$ and $j$ are also connected to one another, forming a closed triangle like that illustrated Figure \ref{triangle}.  We use the term mixed triangle to refer to the case where two of the states ($i$ and $j$ in the illustrative figure) are of the same regime type and fighting one another.  %\footnote{An alternative feature, what we refer to as the multi-network triangle, refers to a process where $k$ is connected to $i$ and $j$ through one network, but $i$ and $j$ are connected through another.  In this case, $i$ and $j$ might not be connected through the MID network, but instead, the alliance network. While this may reflect an ideal theoretical process, these figures, by definition, should be less common than mixed triangles given that they add an additional component to their calculation.  As such, we focus upon the conventional triangle statistic.}
As noted by \citet{cranmer2011inferential}, two-stars are fairly prevalent in the MID network while triangles are exceptionally rare (though they were not considering mixed two-stars or triangles).  We build upon their finding and develop a theory regarding the tie-generating roles of mixed two-stars and mixed triangles.\footnote{This approach stands in contrast to the theoretical approach of \citet{maoz2007enemy} examines different forms of triadic balance in international phenomena.  While finding that states with common enemies tend to ally with one another, they fail to consider the important detail of the regime type of each state.  By accounting for regime type, as \citet{maoz2007enemy} acknowledge may be important, we can rigorously assess the democratic peace and this counter-intuitive finding.}
% R04C01_BC: Added this footnote to Maoz et al 2007 to discuss how we improve

The decision of a state to initiate or join a dispute with another of an opposite regime type may be contingent upon the decisions of other actors of the same regime type.  Consider the decision of the United States to join the United Kingdom in the European theater of World War II. The decision of the U.S. to fight Nazi Germany was conditioned by the U.K.'s prior engagement against Nazi Germany.  When alliances are present, the case is even more compelling. The very nature of military alliances almost necessarily implies that dyadic conflicts are interdependent.  Because states of a similar regime type are more likely to share interests and form alliances, they are also more likely to coordinate on MIDs, regardless of whether the coordinating states are autocratic or democratic.

% I can formally derive this relationship if need be, but this is meant to be more of an analytical example
Consider the costs associated with conflict through a toy model---democracy $i$ is considering fighting autocracy $k$ to resolve an otherwise intractable dispute.  If the cost of the war is smaller than the expected gain of the war (the product of the gain associated with winning the conflict and the probability of winning the conflict), then state $i$ would be expected to initiate the dispute.  If the cost is larger, then state $i$ may seek a partner $j$ to share the burden (direct coordination) or wait for state $j$ to initiate an unrelated dispute with state $k$ that would divert state $k$'s resources (indirect coordination).  For direct coordination, assuming the gain of winning is fixed across the number of direct collaborators $j$, each additional partner decreases the cost of the conflict and increases the probability of winning, increasing the likelihood of the initial conflict or a state joining the conflict.   For indirect coordination, assuming the gain of winning is fixed across the number of indirect collaborators $j$, each additional state $j$ involved in an unrelated dispute with state $k$ increases the probability of states $i$ and $j$ winning by forcing state $k$ to split their resources, increasing the likelihood of the initial conflicts or any other state initiating a dispute.  Having partners, even indirect collaborators, can also increase the perceived international legitimacy of a military action. This toy model is simplistic, but highlights how the decision to engage in a dispute can be interdependent.  This sort of process is common since international crises often receive organized and internationally coordinated responses.

Given that shared interests may be a necessary condition for states to coordinate a MID, it would make sense that likely collaborators would be of a common regime type.  Although, this is not always necessary, interests can be heterogeneous and extend beyond those defined by regime type.  In World War II, the U.S.S.R. certainly collaborated with democracies: the common interest between the U.S., the U.K., and the U.S.S.R to remove Hitler from power lead to their coordination.  It is documented that regime type influences the likelihood that a state joins a conflict, with democracies being among those much more likely to join a dispute as a collaborator \citep{corbetta2005danger}.  Thus, we should expect that states are more likely to join a conflict when their prospective partner is of similar regime type \citep{gartzke1998kant, gartzke2000preferences, werner2000effects, corbetta2005danger}.  Regardless of the severity of the dispute, we expect states to seek out cost-sharing partners.
%While there may be more of an incentive to find conflict partners in fatal MIDs, the costs of additional partners in terms of the division of spoils increases as well.  While incentives persist to share costs for low-level disputes, the potential cost of a reduced share of spoils is less as well.
This leads to the following hypothesis:
	\begin{enumerate}
	\item[$H_2$] Mixed two-stars are a prevalent feature of the conflict network.
	\end{enumerate}

% I can formally derive this relationship if need be, but this is meant to be more of an analytical example
% Because of inefficiencies brought by fighting enemies of enemies, there is no triadic closure, and as such, the democratic peace/autocratic peace persists
% willing to look past small disputes because there are larger concerns
We propose that the democratic and autocratic peaces are a function of the inefficiencies of fighting an enemy of an enemy. Consider the toy model discussed above.  States $i$ and $j$ have a new set of cost calculations they must make should they decide to fight each other-- not only would there be new costs imposed from a new conflict, but the probability of winning the initial conflict decreases as both are mutually undermining the capacity of their partner and splitting its attention and resources across multiple fronts.  One implication of this cost-sharing theory is that states $i$ and $j$ should be less willing to fight each other when they are engaged in mutual conflict against a common enemy. Moreover, this sort of mixed-triadic closure should be less common when conflicts are severe (e.g. involving violence) than when the conflicts in question are lower-level political squabbles because the costs are significantly higher.
%We build upon this by suggesting that $i$ and $j$, while unlikely to be tied through the conflict network, are more likely to be tied through a network highlighting the degree of peacefulness between states.  The theoretical intuition for this comes from \citet{goertz2016puzzle}, who argue that states with relatively peaceful relations come to perceive the world similarly, have common interests, and would be more likely to engage in high-level cooperation.  The existence of a tie within this positive peace network would not only indicate the absence of conflict between two states, but also the ability of the states to engage in warm relations.
This leads to the following hypothesis:
	\begin{enumerate}
	\item[$H_{3}$] Mixed triangles are unlikely in the conflict network.
	\item[$H_{4}$] The probability of mixed triadic closure should be higher in the network of lower-level disputes than the network of serious military conflicts.
	\end{enumerate}

%Once accounting for democratic or autocratic dyads engaged in conflict with the same state, and the tendency for these states to not fight one another due to the inefficiencies of triadic closure, we expect that the typically robust suppressive effect of joint democracy (and joint autocracy) on military conflict will no longer hold. This is because we expect the variance in conflict propensities typically attributed to shared regime type to be explained by the tendency of like-regime states to coordinate against common enemies and for states in conflict with a common enemy not to fight one another.  This leads to our final hypothesis:
%	\begin{enumerate}
%	\item[$H_{5}$] Joint democracy (autocracy) does not have a suppressive effect on conflict after accounting the tendency of states with common enemies not to fight one another.
%	\end{enumerate}

With our theory and hypotheses presented, we now turn to our empirical strategy.

\section{Empirical Strategy}
Our empirical analysis draws on the model of the democratic peace used by \citet{gartzke2013permanent} and improved upon by \citet{dafoe2013democratic}, then extends this specification to include the network effects discussed above and an appropriate inferential strategy for the conflict network.

\subsection{The Temporal Exponential Random Graph Model}
Breaking with much of the conventional conflict literature, which relies upon dyadic regression models to study MIDs, we utilize the temporal extension of the Exponential Random Graph Model (TERGM) in order to conduct valid statistical inference in the presence of network interdependencies.  The strength of the TERGM lies in its ability to simultaneously analyze the effect of monadic and dyadic variables in addition to network dependencies on the initiation of military conflict \citep{cranmer2011inferential, cranmer2016critique}.  The TERGM has been used with high regard to produce sound and meaningful inferences in the study of a variety of international phenomena, including sanctions \citep{cranmer2014reciprocity}, alliance dynamics \citep{cranmer2012complex, li2017three}, and transnational terrorism \citep{desmarais2013forecasting}.
% R02C06_BC: reframed the final sentence of this to note how the TERGM has been seen to produce sound inferences.

The Exponential Random Graph Model (ERGM) and its temporal extension (TERGM) provide an inferential means of estimating the probability of the observed network $N$ conditioned on a vector of specified statistics observed on the network $\bh(N)$.  The model is written as
\begin{equation}
\mathcal{P}(N, \bt) = \frac{\exp \{ \bt' \bh(N)  \}}{\sum_{N^* \in \mathcal{N}} \exp \{\bt ' \bh(N^*)  \}}
\end{equation}

\noindent where $\bh(N)$ refers to the vector of statistics of interest (monadic and dyadic covariates as well as endogenous network structures), $\bt$ refers to the coefficients associated with the prevalence of these statistics, $\exp \{\bt' \bh(N) \}$ refers to a positive weight or the linear predictor, and the denominator $\sum_{N^* \in \mathcal{N}} \exp \{\bt ' \bh(N^*)  \}$ is a normalizing constant.  The normalizing constant can be thought of as the weights (or linear predictors) associated with all possible versions of a network, $\mathcal{N}$, with the same number of nodes as $N$.  This normalizing constant is essential as it allows the analyst to compute the probability of the network $N$ as a function of the covariates whilst scaling it by all the weight of all possible versions of the network such that the quotient fulfills the probabilistic axiom that a probability is $[0,1]$ bound.  This normalizing constant can rarely be estimated directly as $ \mathcal{N}$, all possible permutations of the network, quickly becomes computationally intractably large as the number of actors in the network increases. As such, this is conventionally approximated using Monte Carlo methods or, as we use here, bootstrapped maximum pseudolikelihood (MPLE) for temporal ERGMs \citep{desmarais2010consistent}.\footnote{Some have raised concerns about the reliability of MPLE for ERGM confidence interval estimation \citep{van2009framework}.  However, recent work has indicated that the use of bootstrapped MPLE for TERGM produces both unbiased coefficient estimates and consistent confidence intervals \citep{desmarais2010consistent}.}
% R02C02_BC: Resolved on 8/1 by adding additional detail and discussion.  Reviewer said that the ERGM isn't very clear, and that the tests are difficult to follow.  Added additional detail here.  Will look throughout and revise accordingly.
% R01C01_BC: Resolved on 8/1 using this footnote.  Reviewer 1 expressed concern that MPLE is unreliable, citing Wasserman and Robins 2005 and Corander 1998.  I added a footnote and discussion of Desmarais and Cranmer 2010, 2012.

The TERGM is the empirical model that most closely mirrors our theoretical model.  It allows us to assess longitudinal networks and test the observable implications of our network theory while seamlessly incorporating the nodal, dyadic, and temporal covariates that many have acknowledged in the literature as central to modeling conflict.\footnote{The purpose here is not to make the case of using network models in general, or to compare the TERGM other network models.  This has been done elsewhere \citep{cranmer2017navigating}.  We use the TERGM here as we believe it allows us to best assess our topology-based network theory.}  For a more detailed review of the ERGM and TERGM, we refer the reader to \citet{cranmer2011inferential},  \citet{cranmer2017navigating} and \citet{leifeld2017temporal}.

The results of an ERGM and TERGM can be interpreted at both the network level and the micro-level, much in the same way as many conventional models \citep{desmarais2012micro}.  The effect of a network statistic specified in $\bh(N)$ shares the same substantive sign and significance interpretation as generalized linear models---a positive parameter indicates a positive relationship between the variable of interest and the odds of observing that feature or tie based upon the change statistic associated with an edge.  For example, at the network level, a positive triangle parameter indicates that the likelihood of observing triadic closure within a network is high relative to a random graph with the same number of nodes.  At the micro-level, however, a positive triangle parameter also indicates that a tie between two actors increases as the two actors gain shared partners (i.e., friends of friends) \citep{desmarais2012micro}.  % R01C02_BC: Resolved on 8/1 by expanding this paragraph.  Reviewer expressed confusion over how to interpret ERGMs... They said that the prose implies that the likelihood of conflict between i and j increases because doing so would complete a mixed k-star with k.  But, they said parameters are measured at cross-sectional level so the mixed k-star is an indicator of the total number of these configurations in a given year.  I expanded a discussion that both are okay as there are both network and micro-level interpretations of ERGMs.

\subsection{Data \& Measures}

%---------------------------------------------------------------
% BROUGHT DOWN FROM ABOVE AS IS OPERATIONAL NOT THEORETICAL

Our primary outcome of interest is fatal militarized interstate disputes (FMIDs).
Because the use of deadly force requires, at a minimum, a relatively high level of commitment and risk acceptance  on the part of the initiating government, our theory is most likely to be manifest among such serious conflicts.
Our use of FMIDs is not as a correction to the heterogeneity in MIDs as \citet{hegre2000development} would recommend, but instead reflects a distinct theoretical process.\footnote{\citet{gibler2017heterogeneity} note that the use of FMIDs may underestimate the pacifying effect of joint democracy.  However, the logic of the democratic peace should travel most cleanly to fatal MIDs. In a baseline model specification presented later, joint democracy has a robust effect.}
In particular, we are interested in cases where there are significant costs to MIDs which are typically manifested in the form of human casualties.
%---------------------------------------------------------------

\begin{figure}
\centering
\begin{subfigure}{.5\textwidth}
\hspace{.5cm}
\includegraphics[width=.8\linewidth]{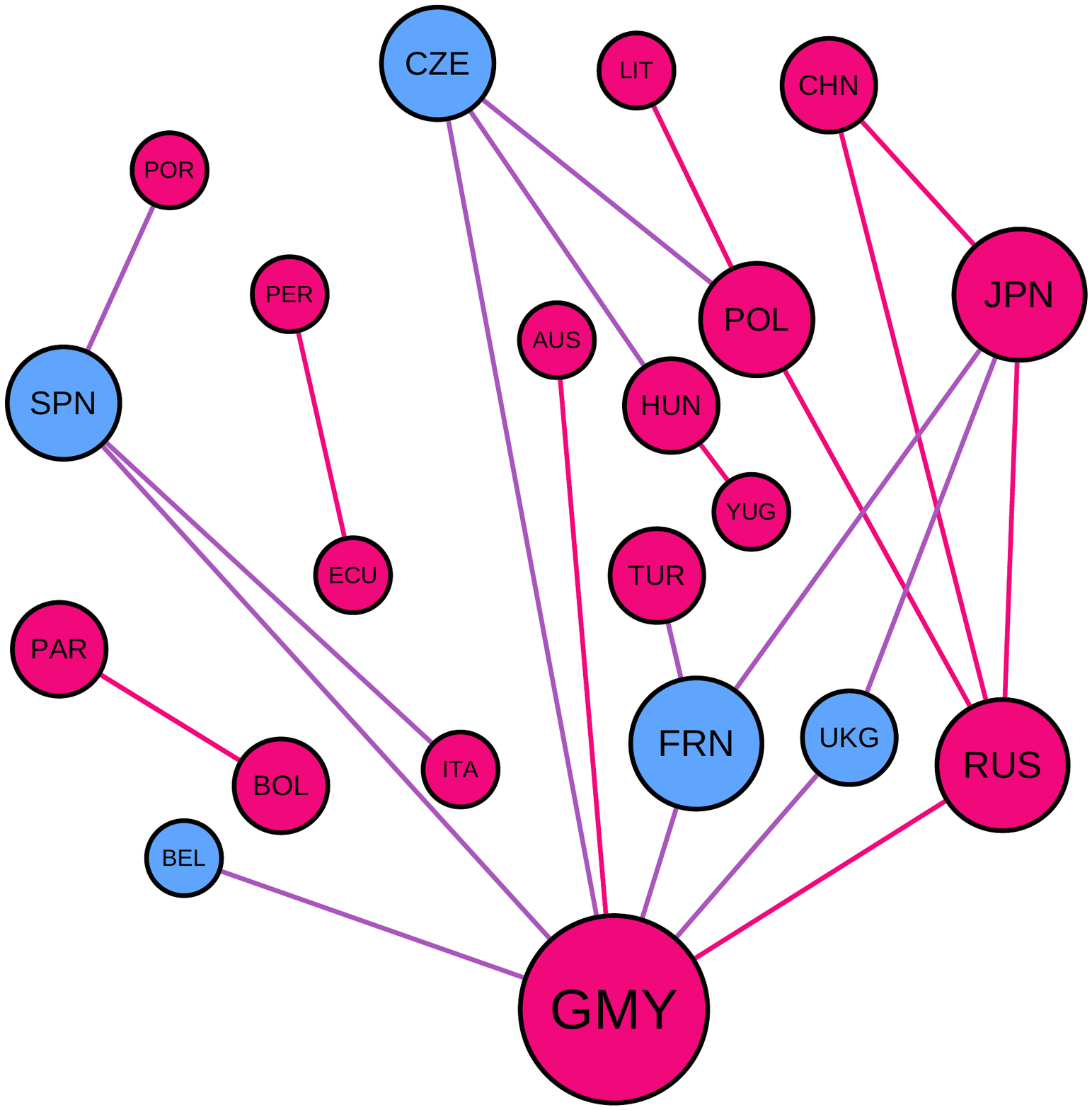}
\caption{Network of all MIDs (1938)}
\label{mids1938}
\end{subfigure}%
\begin{subfigure}{.5\textwidth}
\centering
\includegraphics[width=.8\linewidth]{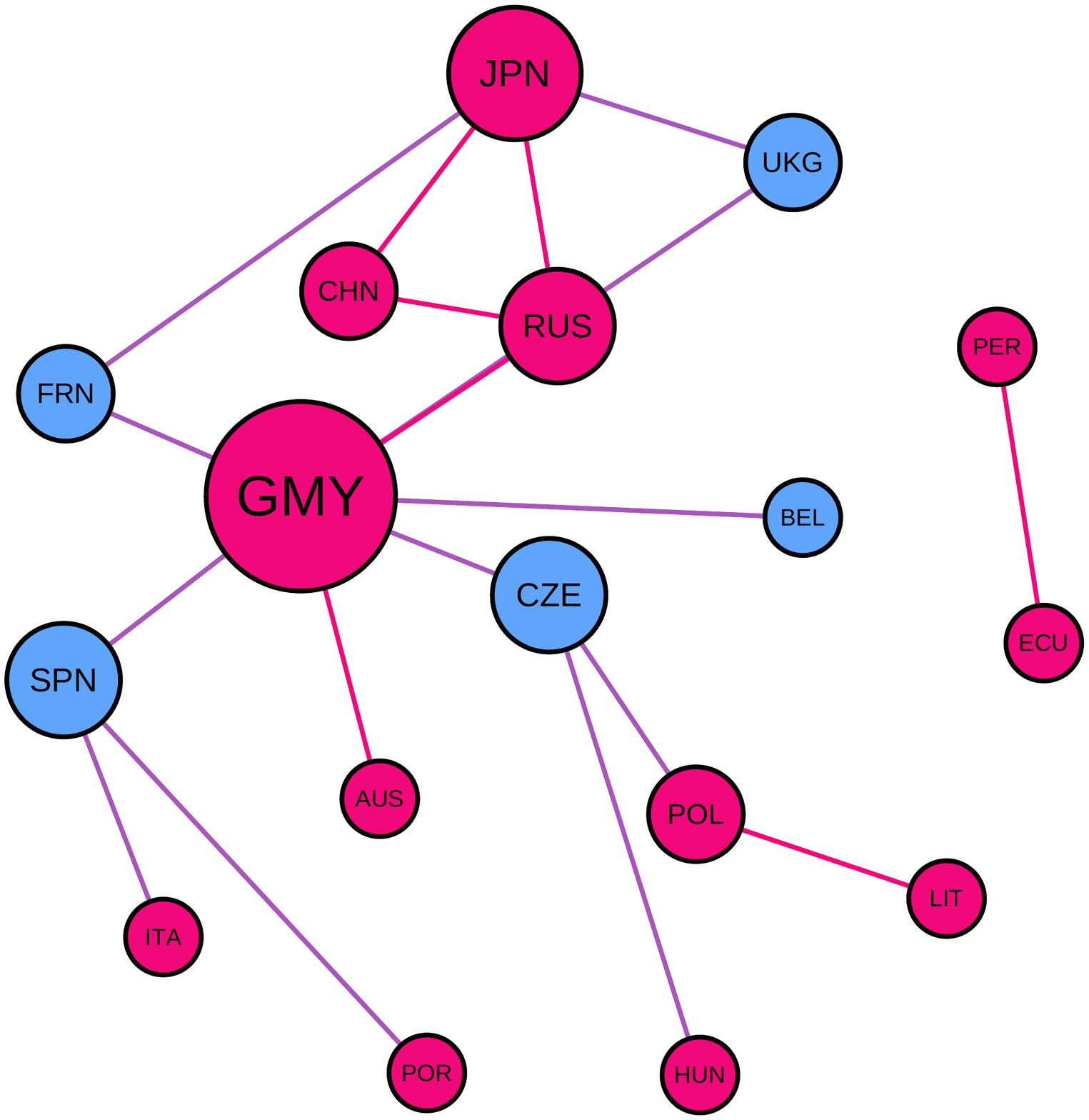}
\caption{Network of all fatal MIDs (1938)}
\label{fatalmids1938}
\end{subfigure}
\caption{\textbf{Networks of All MIDs and FMIDs in an example System Year.}  Democratic states are represented in blue, autocratic states are represented in red.}
\label{netplots}
\end{figure}

To compile our data, we begin from $687,227$ dyad-years reflecting all dyads in the state-system between 1816 and 2001.  These dyad-years were then translated into $186$ undirected binary network-years; a state is included in the network if they were a member of the state system during that year.  From this we form two networks, one of states tied through dyadic engagement in a MID of severity four or greater (FMIDs) and one of states tied through dyadic engagement in a MID regardless of severity (MIDs).  We choose to examine all MIDs/FMIDs and not just originators as it is a more direct test of the cost-sharing portion of the theory.  Figure  \ref{netplots} presents the network of MIDs and FMIDs for 1938.  As is made clear, there is considerable evidence of a ``pile-on'' effect, where many states are engaged in collaborative efforts against common enemies, often of a different regime type.  Naively, there also does not appear to be significant evidence of triadic closure.  %In the coming sections, we will present our measurements for the network dependencies predicted by our theory and exogenous dyadic covariates.

\subsubsection{Network Dependencies}
Given our network theory of conflict, the democratic (or autocratic) peace may be understood as a function of higher-order (indirect) network effects.  In particular, we expect the prevalence of dyadic conflict to be a function of interests and cost-sharing between states with common interests which then goes on to deter states from initiating conflict with collaborators.  As previously discussed, this dynamic may be jointly considered as a function of two network dependencies:  mixed two-stars and mixed triangles.

The mixed two-star is an extension of the conventional two-star to include nodal attributes in the calculation of the statistic.  The mixed two-star is calculated as
\begin{equation}
\bh_{MTS}(N) = \sum_{i > j > k}^{N}(N_{ij}D_{i}A_{j})(N_{jk}A_{j}D_{k}),
\end{equation}

%Conventionally, the number of two-stars within a network is calculated as
%\begin{equation}
%\bh_{TS}(N) = \sum_{i > j > k}^{N}N_{ij}N_{jk},
%\end{equation}

\noindent where the value of $\bh_{MTS}(N) $ is a function of the sum of all instances where a democratic state $i$ is tied to autocratic state $j$  ($N_{ij}$) which is then tied to democratic state $k$ ($N_{jk}$).  $D_{i}$ and $D_{k}$ refer to indicators for whether state $i$ and $k$ are democratic, and $A_{j}$ refers to whether state $j$ is autocratic.  This configuration refers to a coordinated MID where two democracies are ``teaming up" on an autocracy.\footnote{The mixed two-star and mixed triangle statistics, currently unsupported in the \texttt{R} package \texttt{ergm}, had to be developed using the \texttt{ergm.userterms} package \citep{hunter2008ergm}.}   We also account for the complement of this dynamic with autocracies $i$ and $k$ and democracy $j$.

% However, to flip this dynamic, one could easily account for the possibility of two autocrats $i$ and $k$ teaming up on a democracy $j$ by writing the statistic as follows:
%\begin{equation}
%\bh_{MTS}(N) = \sum_{i > j > k}^{N}(N_{ij}A_{i}D_{j})(N_{jk}D_{j}A_{k}).\
%\end{equation}

The mixed triangle is an extension of the conventional triangle statistic widely used in the study of networks.  In particular, it accounts for the presence of actors of different types.  It is calculated similar to the mixed two-star to  account for the possibility of triadic closure within a mixed two-star in the mixed triangle:
\begin{equation}
\bh_{MT}(N) = \sum_{i > j > k}^{N} N_{ik}[(N_{ij}A_{i}D_{j})(N_{jk}D_{j}A_{k}))].
\end{equation}

To be clear, this is expressed similarly to the mixed-two star with the addition of a term accounting for a tie between two nodes of the same type $i$ and $k$ in the same network.
%\noindent where $N_{ij}$, $N_{jk}$, $N_{ik}$ refer to as edges between all nodes included in the subgraph, which produce a product of one for one subgraph, summed over all subgraphs. This general form can be extended to account for the possibility of triadic closure within a mixed two-star in the mix-triangle:
%\begin{equation}
%\bh_{MT}(N) = \sum_{i > j > k}^{N} N_{ik}[(N_{ij}A_{i}D_{j})(N_{jk}D_{j}A_{k}))].
%\end{equation}

%\begin{equation}
%\bh_{T}(N) = \sum_{i > j > k}^{N} N_{ij}N_{jk}N_{ik},
%\end{equation}
%To be clear, this is expressed similarly to the mixed-two star with the addition of a term accounting for a tie between two nodes of the same type $i$ and $k$ in the same network.

To assist with model fit and further appraise theoretical specification, we utilize two other network measures.  First, in each model introduced here we include an ``isolates" term, which is used to measure a network tendency for certain states never to experience a MID in a calendar-year---a striking empirical feature of the sparse conflict network.  Second, for one model specification we replace the mixed-triangle term with a more conventional measure of triadic closure---Geometrically Weighted Edgewise Shared Partners (GWESP).  GWESP allows the user to down-weight the tendency of a network to experience triadic closure to produce the best fitting model.  We use this term as a simpler operationalization of our theory, and robustness test, but it does not account for mixed regime types within those triangles.

\subsubsection{Dyadic Covariates}
In addition to our network statistics, we include a number of dyadic covariates in our model.  While there is certainly not a ``standard" model of conflict, we include many of the factors conventionally thought to influence conflict \citep{gartzke2013permanent, dafoe2013democratic}.  In our primary model, we include nine covariates.  First, and perhaps most centrally, we include a dyadic indicator for whether the lowest POLITY IV within a dyad takes on a value greater than six \citep{jaggers1995tracking, dafoe2013democratic}.  Given that our theory is also a theory of the similar preferences that may emerge from jointly autocratic regime, we include an indicator for whether the greatest POLITY IV value within a dyad is greater than six \citep{dafoe2013democratic}.  We also conduct a robustness check using weak-link regime score, or the lowest POLITY IV value for a dyad.

Contiguity is often included in models as conflict, as states that share a border are more likely to have an opportunity and/or motivation to fight \citep{diehl1985contiguity, stinnett2002correlates}.  This variable is ordinal and describes six increasing levels of proximity.  We choose not to use great-circle distance between national capitals due to large amounts of missing data during our period of observation.  %This is an alternative measure of political relevance and may capture proximity in a way contiguity does not.  This is also thought to confound regime type as regime types cluster, which may then influence the likelihood of conflict \citep{cederman2004conquest}.

We include two measures of the relative power of states in a dyad, as do many studies of the democratic peace \citep{singer1988reconstructing}.
We follow \citet{dafoe2013democratic} and define this ratio as:
\begin{equation}
CR_{AB} = \frac{CINC_{high}}{CINC_{A} + CINC_{B}}
\end{equation}
%One may often use a measure of capabilities that calculates the quotient of the weakest state in a dyad's Composite Index of National Capability (CINC) score and the sum of CINC scores for the dyad.
We also include just the numerator of this term, which \citep{dafoe2013democratic} argue is a more fine-grained measure of the major power indicator used in many studies.

Beyond capabilities, it is also thought that allies are less likely to go to war for a variety of reasons, including the relative costs to security in fighting allies and the increase in information on relative capabilities \citep{bearce2006alliances}.  As such, we include a variable capturing the level of alliance commitment within a dyad coded in ascending level of commitment \citep{gibler2004measuring}.

We include three functions of time to capture temporal dependence that may exist in MIDs between dyads \citep{carter2010back}, including the number of peace years, peace years squared, and peace years cubed.%While it is standard to use the splines-based techniques proposed by \citet{beck1998taking}, the functions proposed by \citet{carter2010back} are less prone to specification error, are simpler to interpret and in many circumstances capture temporal dependence better.

Finally, we include a measure to account for the opportunity for MIDs in a given year.  This scale term is defined as the natural log of the number of states that exist in a given year.  The logic, presented by \citet{raknerud1997hazard}, is that as the number of states in a system increase for a given year, there should also be a proportionate increase in the systemic propensity for conflict.

\section{Results}
 We find support for our theory in both the fatal MIDs and all MIDs networks: mixed two-stars are both positive and statistically significant, mixed triangles are negative and statistically significant, and once both of these network statistics are accounted for, support for the democratic and autocratic peace drops out.  The precise combination of these effects demonstrates that our theory and its underlying logic help to explain the democratic peace.  Beyond simple statistical significance, we find that these variables allow the analyst to accurately forecast conflict.

\begin{figure}
\centering
\begin{subfigure}{.5\textwidth}
\hspace{.5cm}
\includegraphics[width=.8\linewidth]{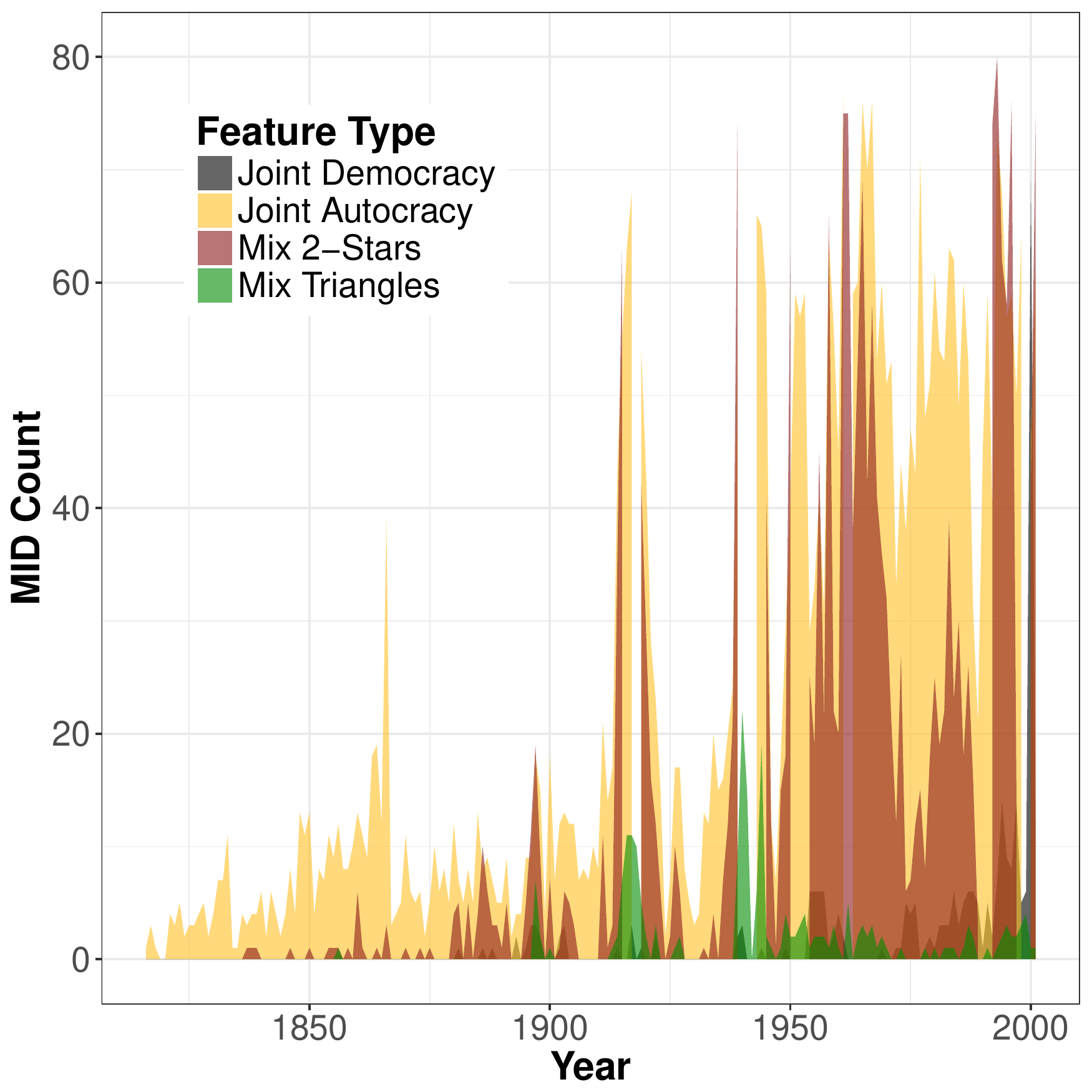}
\caption{All MIDs Network}
\label{rawmids}
\end{subfigure}%
\begin{subfigure}{.5\textwidth}
\centering
\includegraphics[width=.8\linewidth]{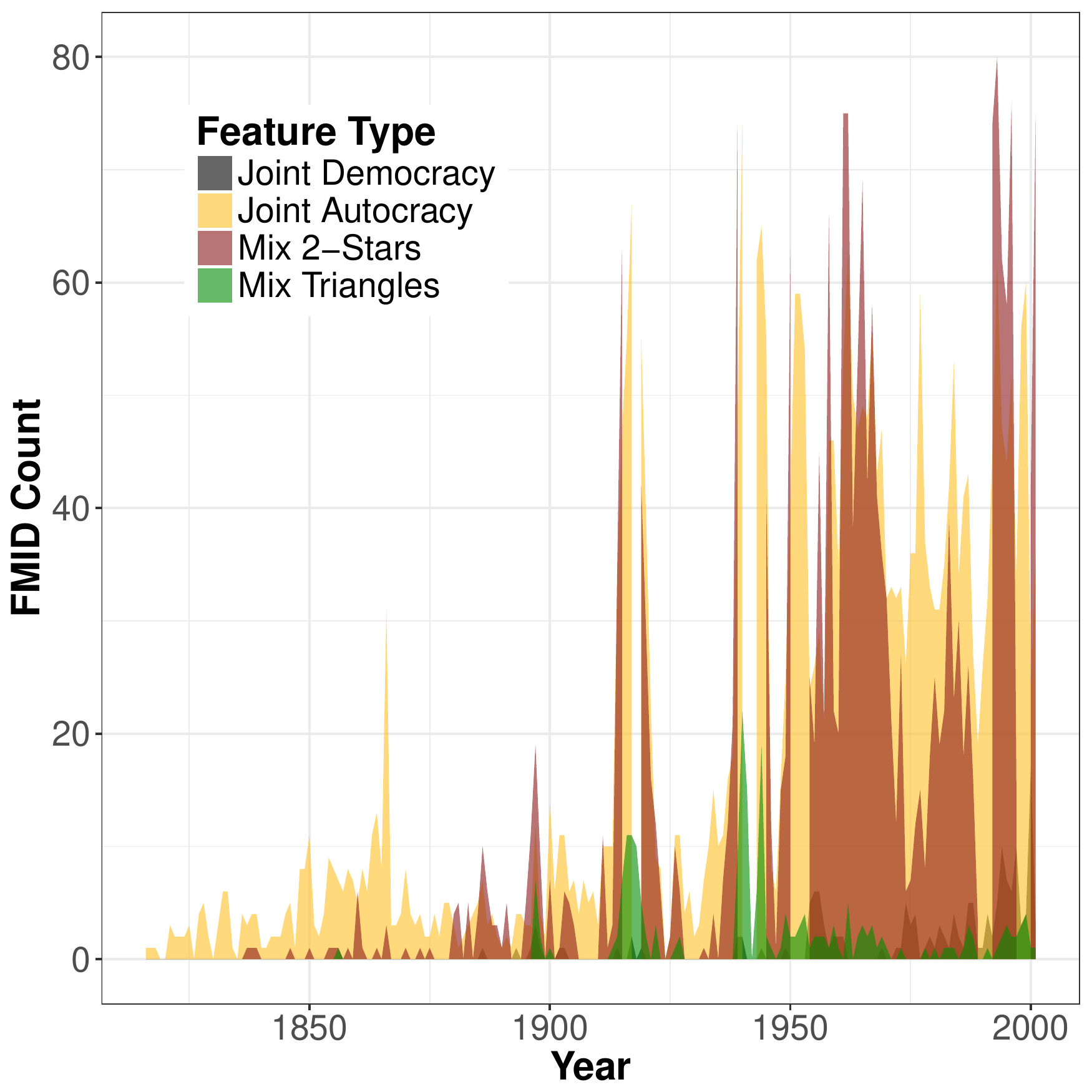}
\caption{All Fatal MIDs Network}
\label{rawfatal}
\end{subfigure}
\caption{\textbf{Raw Count of Network Features Per Year.}  The count of network features are presented per year and colored by feature.}
\label{rawcount}
\end{figure}

The raw counts of each of our primary network dependencies of interest are presented in Figure \ref{rawcount}.  As is made clear, jointly autocratic conflict and mixed two-stars are fairly common features of the MID and fatal MIDs networks.  As is expected, conflict between democracies is fairly rare, as are mixed-triangles.  In this section we attempt to jointly model each of these dependencies to disentangle the confounding effect of mixed two-stars and mixed triangles on the democratic peace and to test our network theory of conflict.

We begin by considering the results of our primary analysis.  We first interpret our results from the models presented in Tables \ref{midModel} and \ref{primaryModel}.  Once our results are presented, we discuss the goodness of fit for these models by assessing how well the model is capable of predicting the last eleven years of conflict.  Then, we present an analysis designed to present the ``smoking gun" for our theory.  We find that the dyads that contribute most to the democratic peace are, in fact, cases of mixed-regime dyads engaging in militarized disputes.\footnote{In the SI Appendix we assess the temporal stability of these effects.}

\subsection{TERGM Results}
Based on our theory, at the dyadic level we expect that mixed dyads would be more likely to engage in conflict.  When considering broader systemic levels, we expect a positive effect for mixed two-stars and a negative effect for mixed triangles.  Tables \ref{midModel} and \ref{primaryModel} support these expectations for the MID and fatal MID networks, respectively.\footnote{Goodness of fit routines that compare observed statistics to networks simulated from the models indicate well, although not perfect, fitting models.  These diagnostics, in addition to other robustness checks, are presented in the SI Appendix.}  In a baseline TERGM omitting our core explanatory variables, Model 1, with all relevant variables save those operationalizing our theory, we uncover a robust democratic and autocratic peace in both the MID and fatal MID networks.%  It appears that fatal and conventional MIDs among jointly democratic and autocratic dyads are relatively rare when excluding relevant network covariates.  Contiguity and CINC score are also in the expected direction, and it appears that at any point in time, it is common for countries to be isolates in the network.

To test our theory, we specify three different models with varying network covariates.  Model 2 reflects a simplified operationalization of our theory using mixed two-stars and as opposed to the mixed triangle, a conventional triangle statistic, Geometrically Weighted Edgewise Shared Partners (GWESP).\footnote{For any model including GWESP we use a geometric downweight of $\alpha = 0.5$ which is indicative of a fairly moderate downweight placed on the prevalence of triangles in the network.  This weight is used to assist with model identifiability.} Our ideal specification is presented in Model 3 for both tables, which includes the baseline specification of Model 1 with mixed two-stars and mixed triangles.  Model 4 builds upon Model 3 by adding weak-link regime score, the lowest Polity regime score within a dyad.\footnote{In Model 4 this variable is added in addition to the dyadic indicators for two reasons.  First, wby including both sets of variables the weak-link regime score should be interpreted as contributing a measure of what happens between the thresholds of the indicators.  Second, the addition of the dyadic indicators improves the fit of the model overall which increases the accuracy of the model parameters underlying the data generating process.  Replacing the joint democracy indicator with weak-link regime score yields the same substantive interpretation.}  Across these specifications, we find evidence supporting our theory and problematizing the democratic peace.

First, a positive coefficient for the ``mixed two-star" terms for both networks implies that mixed dyads are more likely to engage in conventional and fatal MIDs, demonstrating support for $H_{1}$.   In addition, we find support for $H_{2}$,  that mixed two-stars are a prevalent feature of both the MID and fatal MID networks.  Regardless of the model presented in Tables \ref{midModel} or \ref{primaryModel} we find that the mixed two-star term's effect is both positive and statistically reliable.   This demonstrates that mixed-regime dyads are more likely to experience dyadic MIDs or FMIDs than dyads that have a common regime-type.  This may indicate that the democratic peace is a function of mixed dyadic conflict and a conflict of preferences.  Second, and more importantly, the positive and robust effect for mixed two-stars demonstrates that states of a common regime type are likely to explicitly or implicitly coordinate in disputes against states of a different regime type. This supports our cost-sharing model of conflict collaboration whereby states may look for conflict partners with common interests, and in particular, the same regime type.

Additionally, we find support for $H_{3}$ -- that mixed triangles are unlikely in conflict networks. This is demonstrated with a negative and statistically significant effect for both the GWESP effect in Model 2 and the mixed triangles term in Models 3 and 4 as presented in Tables \ref{midModel} and \ref{primaryModel}.  These results support the theoretical claim that triadic closure within conflict networks is costly, especially for states of the same regime type.  This effect, working in conjuncture with the positive effect for mixed-two stars, provides support for our theory overall since our main predictions were positive effects for mixed two-stars and negative effects for mixed-triangles.
\begin{table}
\begin{center}
\scalebox{0.9}{
\begin{tabular}{l c c c c }
\hline
 & Model 1 & Model 2 & Model 3 & Model 4 \\
\hline
Edges                  & $0.12$            & $-0.74$           & $\mathbf{-4.02}$  & $\mathbf{-4.00}$  \\
                       & $[-0.89;\ 1.08]$  & $[-1.54;\ 0.07]$  & $[-5.65;\ -2.81]$ & $[-5.63;\ -2.83]$ \\
Contiguity             & $\mathbf{0.57}$   & $\mathbf{0.59}$   & $\mathbf{0.57}$   & $\mathbf{0.57}$   \\
                       & $[0.53;\ 0.61]$   & $[0.55;\ 0.63]$   & $[0.52;\ 0.62]$   & $[0.52;\ 0.62]$   \\
Capability Ratio & $-0.07$           & $-0.03$           & $-0.19$           & $-0.19$           \\
                       & $[-0.23;\ 0.07]$  & $[-0.19;\ 0.12]$  & $[-0.43;\ 0.02]$  & $[-0.42;\ 0.02]$  \\
CINC (High)            & $\mathbf{6.50}$   & $\mathbf{4.51}$   & $\mathbf{5.18}$   & $\mathbf{5.28}$   \\
                       & $[5.86;\ 7.19]$   & $[3.92;\ 5.16]$   & $[4.29;\ 6.09]$   & $[4.44;\ 6.23]$   \\
Alliance               & $\mathbf{0.07}$   & $\mathbf{0.05}$   & $0.05$            & $0.05$            \\
                       & $[0.02;\ 0.11]$   & $[0.01;\ 0.09]$   & $[-0.02;\ 0.11]$  & $[-0.01;\ 0.11]$  \\
ln States in System    & $\mathbf{-0.46}$  & $\mathbf{-0.50}$  & $\mathbf{-0.46}$  & $\mathbf{-0.45}$  \\
                       & $[-0.63;\ -0.28]$ & $[-0.64;\ -0.34]$ & $[-0.65;\ -0.27]$ & $[-0.64;\ -0.27]$ \\
Peace Years            & $\mathbf{-0.26}$  & $\mathbf{-0.25}$  & $\mathbf{-0.24}$  & $\mathbf{-0.24}$  \\
                       & $[-0.30;\ -0.23]$ & $[-0.28;\ -0.22]$ & $[-0.28;\ -0.21]$ & $[-0.28;\ -0.21]$ \\
Peace Years Squared    & $\mathbf{0.00}$   & $\mathbf{0.00}$   & $\mathbf{0.00}$   & $\mathbf{0.00}$   \\
                       & $[0.00;\ 0.01]$   & $[0.00;\ 0.01]$   & $[0.00;\ 0.01]$   & $[0.00;\ 0.01]$   \\
Peace Years Cubed      & $\mathbf{-0.00}$  & $\mathbf{-0.00}$  & $\mathbf{-0.00}$  & $\mathbf{-0.00}$  \\
                       & $[-0.00;\ -0.00]$ & $[-0.00;\ -0.00]$ & $[-0.00;\ -0.00]$ & $[-0.00;\ -0.00]$ \\
 Isolates               & $\mathbf{1.49}$   & $\mathbf{1.34}$   & $\mathbf{1.41}$   & $\mathbf{1.40}$   \\
                       & $[1.35;\ 1.62]$   & $[1.18;\ 1.50]$   & $[1.23;\ 1.57]$   & $[1.23;\ 1.54]$   \\
\rowcolor{Gray}
Joint Democracy        & $\mathbf{-0.76}$  & $0.22$            & $\mathbf{3.44}$   & $\mathbf{3.17}$   \\
\rowcolor{Gray}
                       & $[-1.14;\ -0.40]$ & $[-0.17;\ 0.59]$  & $[2.81;\ 4.26]$   & $[2.53;\ 4.09]$   \\
\rowcolor{Gray}
Joint Autocracy        & $\mathbf{-0.32}$  & $\mathbf{0.56}$   & $\mathbf{3.83}$   & $\mathbf{3.90}$   \\
  \rowcolor{Gray}
                       & $[-0.41;\ -0.22]$ & $[0.40;\ 0.73]$   & $[3.37;\ 4.56]$   & $[3.42;\ 4.75]$   \\
\rowcolor{Gray}
Weak-Link Regime Score &                   &                   &                   & $\mathbf{0.02}$   \\
  \rowcolor{Gray}
                       &                   &                   &                   & $[0.00;\ 0.05]$   \\
\rowcolor{Gray}
Mix Two-Star           &                   & $\mathbf{0.35}$   & $\mathbf{0.44}$   & $\mathbf{0.44}$   \\
\rowcolor{Gray}
                       &                   & $[0.32;\ 0.38]$   & $[0.36;\ 0.56]$   & $[0.36;\ 0.56]$   \\
\rowcolor{Gray}
GWESP (0.5)            &                   & $\mathbf{-0.06}$  &                   &                   \\
\rowcolor{Gray}
                       &                   & $[-0.13;\ -0.00]$ &                   &                   \\
\rowcolor{Gray}
Mix-Triangle           &                   &                   & $\mathbf{-6.72}$  & $\mathbf{-6.74}$  \\
\rowcolor{Gray}
                       &                   &                   & $[-9.18;\ -5.62]$ & $[-9.54;\ -5.71]$ \\
\hline
Num. obs.              & 614834            & 614834            & 614834            & 614907            \\
\hline
\multicolumn{5}{l}{\scriptsize{}}
\end{tabular}
}
\caption{\textbf{TERGM Results, All MIDs.} Bracketed values indicate 95\% confidence intervals, bolded coefficients have confidence intervals that do not include 0 and are thus statistically reliable at the conventional $0.05$ level.}
\label{midModel}
\end{center}
\end{table}

\begin{table}
\begin{center}
\scalebox{0.9}{
\begin{tabular}{l c c c c }
\hline
 & Model 1 & Model 2 & Model 3 & Model 4 \\
\hline
Edges                  & $0.65$            & $-0.67$           & $\mathbf{-4.07}$  & $\mathbf{-4.06}$  \\
                       & $[-0.66;\ 1.85]$  & $[-1.71;\ 0.49]$  & $[-5.99;\ -2.53]$ & $[-6.02;\ -2.65]$ \\
Contiguity             & $\mathbf{0.58}$   & $\mathbf{0.61}$   & $\mathbf{0.58}$   & $\mathbf{0.58}$   \\
                       & $[0.54;\ 0.62]$   & $[0.57;\ 0.65]$   & $[0.53;\ 0.63]$   & $[0.53;\ 0.63]$   \\
Capability Ratio & $-0.10$           & $-0.05$           & $\mathbf{-0.26}$  & $\mathbf{-0.26}$  \\
                       & $[-0.29;\ 0.07]$  & $[-0.23;\ 0.14]$  & $[-0.55;\ -0.01]$ & $[-0.54;\ -0.01]$ \\
CINC (High)          & $\mathbf{6.12}$   & $\mathbf{4.33}$   & $\mathbf{5.03}$   & $\mathbf{5.15}$   \\
                       & $[5.48;\ 6.87]$   & $[3.67;\ 5.02]$   & $[4.03;\ 6.01]$   & $[4.01;\ 6.21]$   \\
Alliance               & $0.06$            & $0.04$            & $0.05$            & $0.05$            \\
                       & $[-0.00;\ 0.11]$  & $[-0.02;\ 0.09]$  & $[-0.02;\ 0.11]$  & $[-0.01;\ 0.11]$  \\
ln States in System    & $\mathbf{-0.55}$  & $\mathbf{-0.49}$  & $\mathbf{-0.53}$  & $\mathbf{-0.52}$  \\
                       & $[-0.76;\ -0.32]$ & $[-0.69;\ -0.30]$ & $[-0.75;\ -0.29]$ & $[-0.72;\ -0.27]$ \\
Peace Years            & $\mathbf{-0.28}$  & $\mathbf{-0.27}$  & $\mathbf{-0.28}$  & $\mathbf{-0.28}$  \\
                       & $[-0.33;\ -0.25]$ & $[-0.31;\ -0.24]$ & $[-0.33;\ -0.24]$ & $[-0.33;\ -0.24]$ \\
Peace Years Squared    & $\mathbf{0.01}$   & $\mathbf{0.01}$   & $\mathbf{0.01}$   & $\mathbf{0.01}$   \\
                       & $[0.00;\ 0.01]$   & $[0.00;\ 0.01]$   & $[0.00;\ 0.01]$   & $[0.00;\ 0.01]$   \\
Peace Years Cubed      & $\mathbf{-0.00}$  & $\mathbf{-0.00}$  & $\mathbf{-0.00}$  & $\mathbf{-0.00}$  \\
                       & $[-0.00;\ -0.00]$ & $[-0.00;\ -0.00]$ & $[-0.00;\ -0.00]$ & $[-0.00;\ -0.00]$ \\
Isolates               & $\mathbf{1.58}$   & $\mathbf{1.41}$   & $\mathbf{1.51}$   & $\mathbf{1.50}$   \\
                       & $[1.42;\ 1.74]$   & $[1.23;\ 1.59]$   & $[1.32;\ 1.67]$   & $[1.31;\ 1.68]$   \\
\rowcolor{Gray}
Joint Democracy        & $\mathbf{-0.91}$  & $0.05$            & $\mathbf{3.83}$   & $\mathbf{3.52}$   \\
\rowcolor{Gray}
                       & $[-1.21;\ -0.63]$ & $[-0.25;\ 0.32]$  & $[3.23;\ 4.79]$   & $[2.82;\ 4.62]$   \\
\rowcolor{Gray}
Joint Autocracy        & $\mathbf{-0.27}$  & $\mathbf{0.60}$   & $\mathbf{4.39}$   & $\mathbf{4.49}$   \\
  \rowcolor{Gray}
                       & $[-0.38;\ -0.17]$ & $[0.42;\ 0.77]$   & $[3.85;\ 5.22]$   & $[3.94;\ 5.47]$   \\
\rowcolor{Gray}
Weak-Link Regime Score &                   &                   &                   & $\mathbf{0.03}$   \\
\rowcolor{Gray}
                       &                   &                   &                   & $[0.00;\ 0.05]$   \\
\rowcolor{Gray}
Mix Two-Star           &                   & $\mathbf{0.39}$   & $\mathbf{0.60}$   & $\mathbf{0.61}$   \\
 \rowcolor{Gray}
                       &                   & $[0.35;\ 0.43]$   & $[0.49;\ 0.72]$   & $[0.49;\ 0.73]$   \\
\rowcolor{Gray}
GWESP (0.5)            &                   & $\mathbf{-0.11}$  &                   &                   \\
 \rowcolor{Gray}
                       &                   & $[-0.21;\ -0.03]$ &                   &                   \\
\rowcolor{Gray}
Mix-Triangle           &                   &                   & $\mathbf{-6.86}$  & $\mathbf{-6.88}$  \\
  \rowcolor{Gray}
                       &                   &                   & $[-8.69;\ -5.96]$ & $[-9.05;\ -5.93]$ \\
\hline
Num. obs.              & 614777            & 614777            & 614777            & 614902           \\ \hline
\multicolumn{5}{l}{\scriptsize{}}
\end{tabular}
}
\caption{\textbf{TERGM Results, Fatal MIDs.} Bracketed values indicate 95\% confidence intervals, bolded coefficients have confidence intervals that do not include 0 and are thus statistically reliable at the conventional $0.05$ level.}
\label{primaryModel}
\end{center}
\end{table}

\begin{figure}[ht]
\centering
\includegraphics[width=0.9\linewidth]{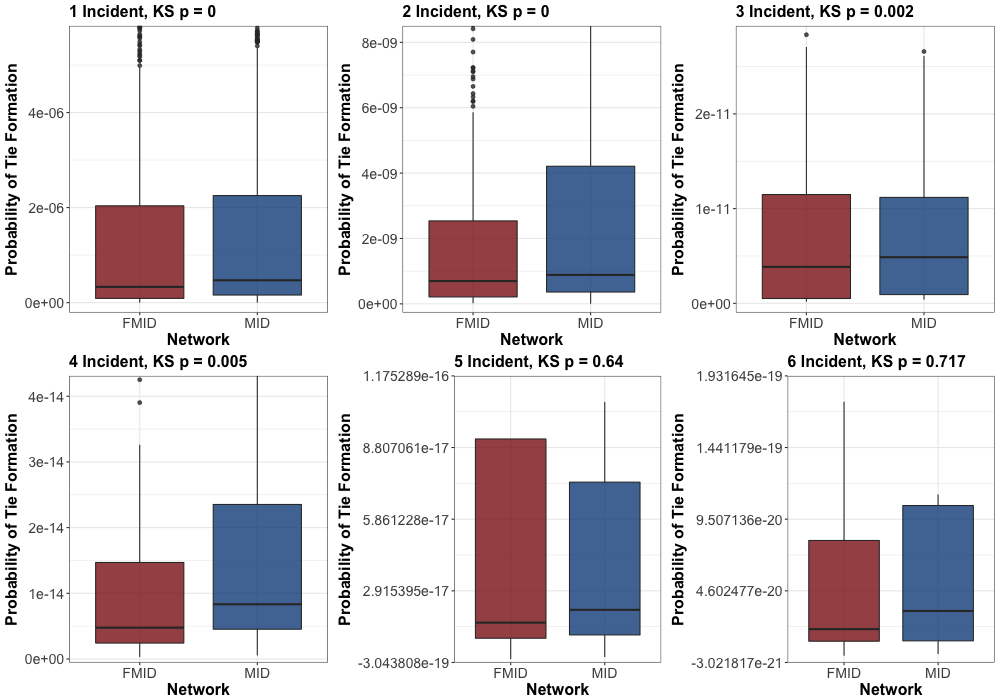}
\caption{\textbf{Probabilities of Triadic Closure.} Plots are stratified by the number of potential mixed-triangled incident upon the edge, boxes are colored by the network.  A pooled Kolmogorov-Smirnov test returns a p-value less than $2.2^{-16}$.}
\label{pprobs}
\end{figure}

Assessing support for $H_{4}$, that aversion to mixed triangles should be a stronger generative feature of the fatal MID network than the all MID network, is difficult as the fatal MID network is simply a thresholded and nested version of the MID network.  By definition, mixed-triangles must be less prevalent in the fatal MID network than the MID network.  To work around this problem, %and assess whether the probability of mixed triadic closure is higher in the MID network than the FMID network
we examine the probability of tie formation between democratic or autocratic states $i$ and $j$ conditional on their shared conflict with an autocratic or democratic state $k$.  If the distribution of these probabilities is higher in the MID network than the fatal MID network then it would indicate that the MID network has a greater tendency towards triadic closure, offering additional support for our theory.  We disaggregate these distributions by the number of potential mixed-triangles incident upon the edge.  Our theory would expect that the cost of mixed-triadic closure increases as the number of shared enemies between states increases.  Figure \ref{pprobs} presents the results from this analysis which confirm our theoretical expectation.  First, for each of the plots presented the probability of triadic closure is higher in the MID network than the FMID network, indicating that the potential costs of triadic closure are much higher as conflict intensity increases.  Second, the probability of tie formation decreases based upon the number of mixed-triangles incident upon an edge, indicating that the costs of fighting a conflict partner increase as they have more common enemies.

To more systematically describe these distributions and their relative locations we utilize two-sample Kolmogorov-Smirnov tests.  A low p-value would indicate that the distribution of probabilities of triadic closure is higher in the MID network than the fatal MID network.  We perform this test for each distribution of mixed-triangles incident upon an edge.  The results of these tests are presented in the sub-figure titles for Figure \ref{pprobs}.  We reject the null hypothesis for four of six tests indicating mixed support for $H_{4}$.  It appears that when more than 4 mixed-triangles are incident upon the edge the probability of triadic closure in the MID network is not significantly higher than that in the FMID network.  When examining the results of a test that pools across these six conditions, a Komogorov-Smirnov test returns a p-value significant at any conventional threshold, indicating support for $H_{4}$.

In addition, once accounting for both mixed two-stars and mixed triangles, the theoretical process we believe to underlie the democratic peace, we find that while the 95\% confidence intervals for the joint democracy and autocracy terms \textit{do not include zero, the signs flip}.  This is presented by Models 3 and 4 in Tables \ref{midModel} or \ref{primaryModel}.  This would indicate that once accounting for common enemies and the inefficiencies of fighting enemies of enemies, purely dyadic conflict between joint democracies or autocracies is actually fairly common. There are 177 fatal MIDs between democracies between 1816 and 2001. Since 1974 there have been 12 disputes between the U.S. and Canada.  As recently as 1991, there was a high profile fatal dispute between the U.S. and Canada over the Northwest Passage. This is just one instance among many that conflict scholars often overlook.  This demonstrates that not only may our theory explain the democratic finding that has been standard in the peace science literature for decades, but that decades of peace science literature on the democratic peace may actually have gotten the answer exactly wrong.  This result may seem counterintuitive to many, but when interpreting the effect in the context of the model, it may seem more intuitive.  The positive effect for the joint democracy and joint autocracy term indicates that when democracies or autocracies do not have common enemies, are not contiguous or allies, when the probability that any given country wins or the capabilities of the strongest state in a dyad are low, jointly democratic or autocratic dyads are more dispute prone relative to mixed dyads.
	% R02C04_BC Added a discussion of the joinly democratic mids to the prior paragraph.

%%% pick up here -- put these in the SI
%While these models do not fit perfectly, overall, these models do appear to fit reasonably well.  Goodness of fit box-plots are presented in Figures \ref{gofReduced} and \ref{gofFull} for Models 1 and 3 respectively.  A model fits best if the black line runs through the median line for each box.  The networks simulated from the parameters estimated appear to closely resemble the observed networks.

%\begin{figure}[ht]
%\centering
%\includegraphics[width=0.75\linewidth]{gof_manuscript_model1_08082017.pdf}
%\caption{\textbf{Goodness of Fit for Baseline Model Specification (Model 1).} 10 networks simulated per time-step.}
%\label{gofReduced}
%\end{figure}

%\begin{figure}[ht]
%\centering
%\includegraphics[width=0.75\linewidth]{gof_manuscript_model3_08082017.pdf}
%\caption{\textbf{Goodness of Fit for Full Model Specification (Model 3).} 10 networks simulated per time-step.}
%\label{gofFull}
%\end{figure}

\subsection{Out-of-Sample Prediction}
In any study of conflict, one exceptionally relevant question is how well the model fits out-of-sample  \citep{ward2010perils, cranmer2017can}.  To evaluate how well our model can forecast fatal MIDs, we partition the data into two datasets.  First, we estimate a baseline model and a full network model with the same specification for Model 3 in Table \ref{primaryModel} on a ``training" dataset that includes all years 1816-1990.  Once training the model we evaluate how well the model can predict the next 11 years, 1991-2001.\footnote{The training model results closely approximate those presented in Table \ref{primaryModel}.}  Receiver Operating Characteristic (ROC) and Precision-Recall (PR) Curves indicate that the model fits the test data quite well.  The network model fits much better than the baseline model without our network measures of interest for all test networks according to area under the ROC Curve (ROC AUC).  This indicates that our network terms assist in avoiding false positives while providing information that allows for the detection of true positives.   Relying upon ROC for rare events data, such as these however, does not give a clear picture as it is fairly easy to predict negatives, or cases of non-conflict, while avoiding false positives.  As such, PR is often used for classification in class imbalanced data as it penalizes for both false positives (predicted conflict where there was no conflict) and false negatives (predicted peace where there was conflict) \citep{cranmer2017can}.  When examining PR, the network model outperforms the baseline model for 6 of 11 test networks.  In other words, the network model outperforms the baseline model on average when managing the trade-off of avoiding false positives and false negatives.
\begin{figure}
\centering
\begin{subfigure}{.5\textwidth}
\includegraphics[width=\linewidth]{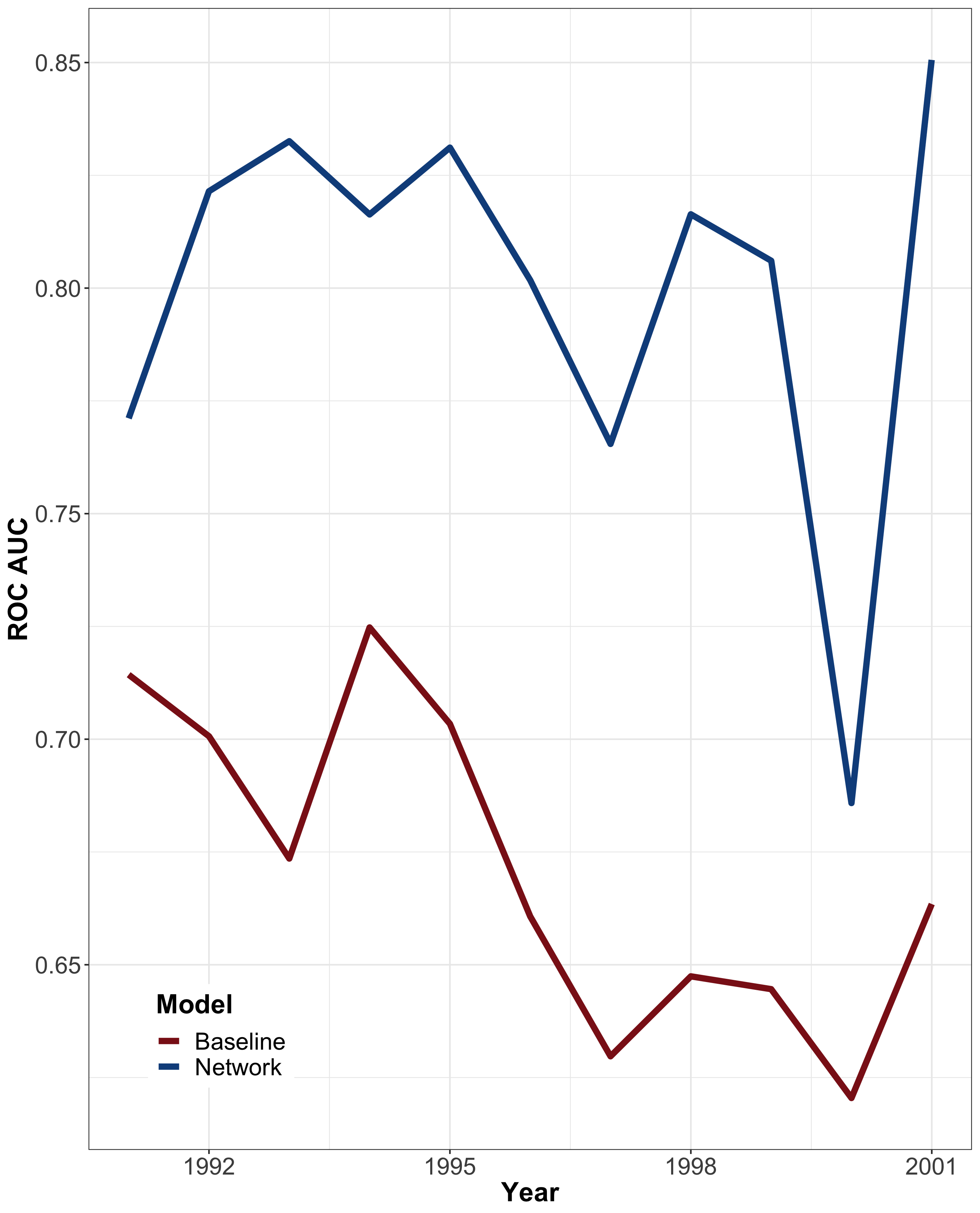}
\caption{ROC Area Under the Curve}
\label{rpc}
\end{subfigure}%
\begin{subfigure}{.5\textwidth}
\centering
\includegraphics[width=\linewidth]{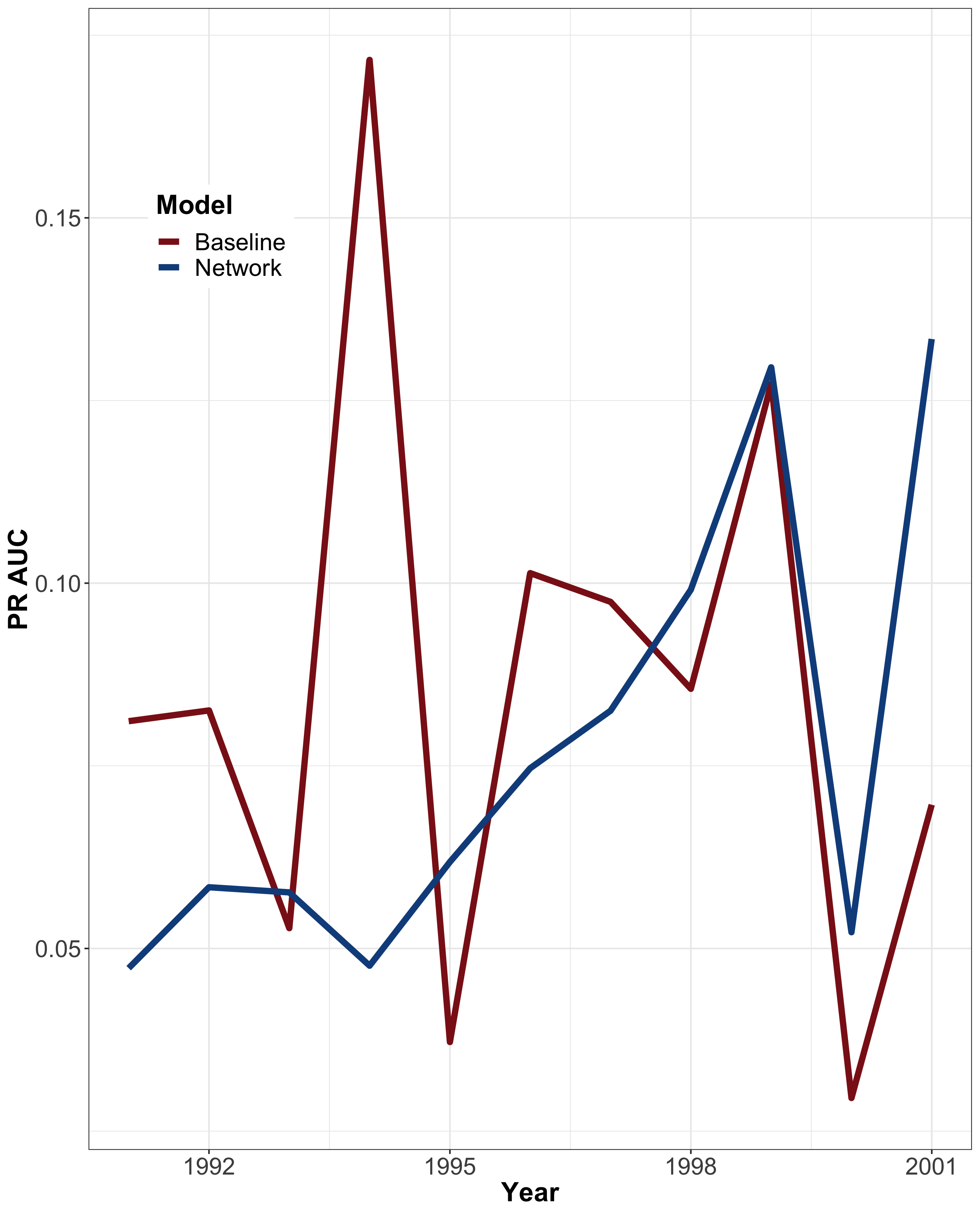}
\caption{PR Area Under the Curve}
\label{pr}
\end{subfigure}
\caption{\textbf{ROC and PR Area Under the Curve by Test Network.}  Network model is colored in blue relative to the baseline model with an isolates term presented.}
\label{subgraphs}
\end{figure}

\subsection{Influential Dyads}
To get a sense of which dyads are contributing the most to the democratic peace, and get a more fine-grained read on our theory, we estimated 19,856 different models iteratively excluding a unique dyad from the estimation procedure.  To estimate this model we use a logistic regression model of fatal MID initiators that included the following variables: Contiguity, Distance (ln), Probability of Winning, Largest CINC Scores, Allies, States in System, Peace Years, Time Functions, and Weak Link Regime Score.  In this case, we transition from the prior dichotomized measure which is useful in the network context to the commonly used weak link regime score, which is the lowest regime score within a dyad.  The following dyads, once removed, lead to the largest increases (or, moving away from large negative values) in the democratic peace coefficient:
\begin{enumerate}
\item U.S. \& North Korea (Mixed Dyad, 54 Years)
\item U.S. \& Canada (Joint Democracy, 82 Years)
\item Iran \& Iraq (Joint Autocracy, 70 Years)
\item U.S. \& China (Mixed Dyad, 142 Years)
\item North Korea \& South Korea (Joint Autocracy, 38 Years; Mixed Dyad, 15 Years)
\item U.S. \& Iraq (Mixed Dyad, 70 Year)
\item Egypt \& Saudi Arabia (Joint Autocracy, 65 Years)
\item U.K. \& Iraq (Mixed Dyad, 70 Year)
\item Iraq \& Kuwait (Joint Autocracy, 41 Years)
\item U.K. \& France (Joint Democracy, 105 Years; Joint Autocracy, 58 Years; Mixed Dyad, 22 Years)
\end{enumerate}

Of the top ten, only a sliver of the dyads most influential for the democratic peace are joint democracies.  Of these ten only two dyads are consistently democratic, including the U.S. \& Canada and U.K. \& France.  Of the remaining cases four were typically mixed dyads  and four were typically jointly autocratic.  We consider these cases the ``smoking gun"---it appears that the democratic peace may be influenced more by conflict between mixed dyads and joint autocracies than by peace among democracies.   The mixed dyads that exercise the most influence over the democratic peace are enduring rivalries.  Of the jointly autocratic dyads, there are many cases where they directly fight one another regularly.  Of the two jointly democratic cases, there are few cases where crises of interests emerge. but many cases of collaboration against a common autocratic enemy.  Consider the Anglo-French dyad following the Entente Cordiale in 1904.  While there were common interests that may have prevented conflict independently, the dyad coordinated to counter many external threats, including the Triple Alliance and Axis Powers.  While the United States and Canada certainly had disputes in the 19th century, peace was maintained in the 20th century due to common interest and coordination in MIDs against common enemies including Kim Il-Sung, Slobodan Milo\v{s}ovi\'{c}, and Saddam Hussein.  During World War 1, World War 2, Kosovo and the Gulf War, it would make little sense for France and Britain or the U.S. and Canada to engage in a MID.

\section{Conclusion}
The dyadic understanding of the democratic peace has become ubiquitous in International Relations.  By looking beyond simple dyadic analysis, accounting for the embededness of states in a much more complex network, we found the democratic peace may not be as robust as previously thought. Our results demonstrate that after accounting for the tendency for like-regime states with common enemies not to fight one another, the effect of the democratic peace not only vanishes, but jointly democratic dyads seem to be \emph{more} conflict prone than mixed dyads. These results are consistent across operationalizations of the outcome variable, our triadic closure predictor, measurements of joint democracy, and a variety of other factors.

We believe this explanation for the democratic peace is not a mechanism for understanding the democratic peace, but instead, an alternative.  What we have shown here is that conflict between democracies indeed exists and the peaceful relations occasionally found are not necessarily a function of the affinity of democratic states, or intrinsic attributes of democratic states, but instead, a function of the strategic inefficiencies of fighting a state with a shared enemy.  While regime type may influence the interests of states, we find that it does not directly influence the probability that any two states fight one another.

%We suspect that these dynamics are informed through the process by which disputes emerge, escalate, and then subsequently pull in other states.  We posit that militarized disputes are most likely to emerge out of conflicting preferences, and these conflicting preferences are informed by the regime types of states.  The most intractable disputes are likely to emerge from these conflicting preferences as they often inform the identity of the state and become existential issues.  From this, states considering disputes may seek out collaborators as to increase their probability of winning and decrease the risk, even if it means the return from conflict may decrease.  Given that fighting the enemy of an enemy decreases the probability of winning and creates a new source of cost, we do not expect these states to fight.  Given our results, we can say that we find support for our theory.  In addition, it appears that these conclusions are consistent across time.

There are three major implications to our research.  First, scholars should be hesitant to consider dyadic conflict in isolation, as there are network dependencies informing whether a state engages or joins a MID.  Second, preferences operating in addition to network interdependencies and collaboration explain much of the democratic peace.  Third, when studying conflict, scholars and practitioners should consider the cost structure of collaboration, and how these dynamics inform not only conflict initiation, but conflict escalation.

Particularly interesting is  that the theoretical mechanism at work here is dramatically simpler than any of the established justifications for the democratic peace. We do not rely on arguments about institutions or norms, but just the simple and intuitive proposition that it does not make much sense for two states fighting a third to also fight each other. What the existing literature seems to have missed, usually theoretically and almost always empirically, is that dyadic conflicts do not occur in isolation, but in the context of a complex network of relations.

% Bibliography

%\bibliographystyle{apsr}
%\bibliography{NetworkDemocraticPeaceBib.bib}
\end{document}